\documentclass[conference]{IEEEtran}

\IEEEoverridecommandlockouts

\usepackage{graphicx,cite,calc,xcolor,subfigmat,amssymb,amsmath,mathrsfs,dsfont,hyperref,epstopdf}
\usepackage[normalem]{ulem}

\usepackage{float}
\usepackage{graphicx}
\usepackage{lipsum}
\usepackage{soul} 
\usepackage{xcolor}
\usepackage{optidef}
\usepackage{listings}
\usepackage{enumitem}
\usepackage{hyperref}
\usepackage{amsthm}
\usepackage[T1]{fontenc}
\usepackage{subfigure}
\usepackage{caption}

\usepackage{algorithm}
\makeatletter
\newcommand*{\rom}[1]{\expandafter\@slowromancap\romannumeral #1@}
\makeatother
\usepackage[normalem]{ulem}
\usepackage{algorithm}
\usepackage[switch]{lineno}
\usepackage {amsmath}
\makeatletter
\newtheorem{proposition}{Proposition}

\usepackage{newtxtext,newtxmath}

\begin{document}

\title{MIMO-NOMA Enabled Sectorized Cylindrical Massive Antenna Array for HAPS with Spatially Correlated Channels}

\author{\IEEEauthorblockN{Rozita Shafie\IEEEauthorrefmark{1}\IEEEauthorrefmark{2},
Mohammad Javad Omidi\IEEEauthorrefmark{2}\IEEEauthorrefmark{3},
Omid Abbasi\IEEEauthorrefmark{1},
Halim Yanikomeroglu\IEEEauthorrefmark{1}
}

\IEEEauthorblockA{\IEEEauthorrefmark{1}Non-Terrestrial Networks (NTN) Lab, Department of Systems and Computer Engineering, Carleton University, Ottawa, Canada\\
\IEEEauthorrefmark{2}Department of Electrical and Computer Engineering, Isfahan University of Technology, Isfahan 84156-83111, Iran\\
\IEEEauthorrefmark{3}Department of Electronics and Communication Engineering, Kuwait College of Science and Technology, Doha 35003, Kuwait\\
Email: \IEEEauthorrefmark{1}rozitashafie@sce.carleton.ca,
\IEEEauthorrefmark{2}omidi@iut.ac.ir,
\IEEEauthorrefmark{1}omidabbasi@sce.carleton.ca,
\IEEEauthorrefmark{1}halim@sce.carleton.ca
}
\thanks{A preliminary version of this work was presented in IEEE Globecom 2022 Workshops \cite{10008738}. }}

\maketitle

\begin{abstract}
The high altitude platform station (HAPS) technology is garnering significant interest as a viable technology for serving as base stations in communication networks.  However, HAPS faces the challenge of high spatial correlation among adjacent users’ channel gains which is due to the dominant line-of-sight (LoS) path between HAPS and terrestrial users. Furthermore, there is a spatial correlation among antenna elements of HAPS that depends on the propagation environment and the distance between elements of the antenna array. This paper presents an antenna architecture for HAPS and considers the mentioned issues by characterizing the channel gain and the spatial correlation matrix of the HAPS.
We propose a cylindrical antenna for HAPS that utilizes vertical uniform linear array (ULA) sectors. Moreover, to address the issue of high spatial correlation among users, the non-orthogonal multiple access (NOMA) clustering method is proposed. An algorithm is also developed to allocate power among
users to maximize both spectral efficiency and energy efficiency while meeting quality of service (QoS) and successive interference cancellation (SIC) conditions. Finally, simulation results indicate that the spatial correlation has a significant
impact on spectral efficiency and energy efficiency in multiple antenna HAPS systems.
\end{abstract}


\section{Introduction}
\subsection{Background}
The sixth generation (6G) of wireless technology aims to achieve higher data rates and broader coverage areas. One promising approach to accomplish this goal is through the use of high-altitude platform station (HAPS) systems \cite{9380673}. 
HAPS technology offers several advantages over traditional satellite and terrestrial communication layers. HAPS systems benefit from favorable channel conditions and geostationary positions, have a smaller footprint compared to satellite nodes, and can provide lower latency \cite{abbasi2023haps_mag}. Moreover, HAPS technology can serve as a super macro base station (SMBS) to cover a large metropolitan area \cite{ARUM2020232,9356529,Omid_hemi}.

It is expected that a massive multiple-input multiple-output (MIMO)  HAPS system will provide significant improvements in both service capacity and area coverage by utilizing favorable antenna arrays \cite{9449056}.
There are various antenna architectures that can be used in HAPS systems, including uniform planar array (UPA), uniform linear array (ULA), and cylindrical antennas \cite{9449056}. 
Most of the current research on massive-MIMO for HAPS systems utilizes the UPA antenna mounted on the bottom of the aircraft facing the earth \cite{10008738}. However, this configuration leads to limited coverage, with a coverage radius of only $ 20-60 \; \rm{km} $ \cite{9449056}.
Recent studies by Softbank and HAPSMobile have shown that cylinder-shaped and large-scale array antennas can provide a higher system capacity and coverage radius \cite{9779715}.
Nonetheless, an important challenge facing the HAPS cylindrical antenna is the high spatial correlation between the antenna elements, which makes the detection of users' signals difficult \cite{8861014}, \cite{7052134}. \par
The issue of spatial correlation in HAPS is influenced by various aspects. In previous research, as highlighted by \cite{10008738},  significant attention has been given to analyzing correlated channels specifically for horizontal UPAs.
In \cite{10008738}, it is observed that the spatial correlation escalates proportionally with the number of antenna elements present on the horizontal UPA.  Recognizing this challenge, employing a large number of antenna elements becomes impractical for HAPS. 
Consequently, the horizontal UPA was limited to 4 and 6 elements, as exceeding this range is deemed impractical for the HAPS BS.
As discussed in the following sections,
we tackle this main issue by opting for multiple sectorized antenna arrays instead of a single large antenna array. This change is designed to minimize spatial correlation among antenna elements.

The spatial correlation in cylindrical HAPS depends on several factors, including the size and shape of the cylinder, the altitude of the HAPS, and the distribution of scatterers in the environment\cite{SIG-093}. 
This spatial correlation is caused by the low vertical angular spread among user equipment (UEs). In the azimuth domain, it is easier to separate UEs than in the elevation domain \cite{SIG-093}.
The propagation environment also impacts the spatial correlations among the antenna elements on the BS.
The difference between a rich scattering environment and a sparse scattering environment in the context of spatial correlation in HAPS systems refers to the degree of diversity in the radio wave propagation environment.
A sparse scattering environment is characterized by a small number of independently fading paths between the HAPS and UE, resulting in a high degree of spatial correlation between the antennas, since the radio waves arriving at different antennas are similar. This high spatial correlation between the antennas limits the degree of freedom in the system, and can consequently restrict the capacity of the communication link. 

\subsection{Related works}
The literature review, as detailed in the studies referenced by \cite{9772280} and \cite{10008738}, highlights the role of HAPS as super macro base stations that can provide connectivity for a variety of wireless access applications.
To enhance coverage and capacity,  a massive MIMO system must be implemented, which can improve diversity gain, SINR, and spatial multiplexing.
However, classical MIMO cannot be directly applied to HAPS systems since the line-of-sight (LoS) links between  HAPS and users create a significant correlation between the channel gain of users, leading to an unfavorable propagation environment as discussed in \cite{7841647} and \cite{9222215}.

In \cite{9449056},  a cylindrical massive MIMO system was proposed to generate multiple high-gain beams for achieving ultra-wide coverage of $100 ~\rm{km}$, and high capacity. This study introduces a beamforming method for cylindrical antenna structures in cells with a large radius, which improves the SINR of area-edge users compared to m-MIMO systems with planar arrays \cite{8891546}.
HAPS can cover a large area and enable direct communication with mobile terminals on the ground.  However, to accommodate high communication traffic, the HAPS system must support multiple cells \cite{patent_key}. This can result in frequent handovers and degradation of communication quality due to the displacement of the coverage area (footprint) on the ground as the platform moves \cite{8746491}, \cite{9178753}. In other words,  the communication area footprint to the ground cannot be fixed due to the aircraft's rotations. To solve this problem, a footprint fixation control method was proposed using a cylindrical array antenna \cite{9348661},\cite{knuthwebsite}.  This cylindrical antenna can control the beam three-dimensionally in any desired direction \cite{knuthwebsite}. 

According to  \cite{8861014},  when there is not much scattering in the propagation environment, the signal distribution is not uniform, and channel gains are spatially correlated. 
Meanwhile, \cite{8620255} considers the signal processing and channel modeling for Rayleigh and Rician fading channels.
The authors of \cite{8861014} suggest that exploiting the natural spatial correlation of propagation channels can resolve the pilot contamination phenomenon, which was initially believed to create capacity limits.
Moreover, \cite{8620255} and \cite{7172496} stress the importance of considering spatial channel correlation in massive MIMO networks.

Power domain non-orthogonal multiple access (NOMA) is an emerging technology for improving the spectral efficiency and energy efficiency of wireless communication systems. 
In NOMA scenarios, users within the same cluster receive the same signal from the BS in downlink mode.  We can solve the problem of spatial correlation between adjacent users by using a NOMA scenario.
\cite{9226489} focuses on developing an energy-efficient algorithm for beamspace HAPS-NOMA systems over Rician fading channels. The authors' proposed algorithm uses NOMA to serve users in each beam simultaneously and obtains the digital precoder by zero forcing (ZF) algorithm based on the equivalent channel.
In \cite{9942358}, the focus is on utilizing NOMA technology in HAPS communications to achieve the connectivity, reliability, and high-data-rate requirements. The authors introduce a user selection and correlation-based user pairing algorithm and implement the polar-cap codebook specifically tailored for NOMA-based HAPS systems.
Also, \cite{8653850}  presents user grouping and beamforming for the massive MIMO HAPS systems. The authors proposed a user grouping scheme based on the average chordal distance between the statistical eigenmodes. 
In \cite{10333511}, the use of multi-beam transmission in HAPS is advocated for high throughput, with each beam dedicated to a cell accommodating multiple users. To address challenges arising from intra-beam and inter-beam interference, the paper introduces an enhanced NOMA scheme.
Authors in \cite{10330559} investigates power allocation and system performance in NOMA-enabled integrated satellite-HAPS-terrestrial systems. Considering practical constraints such as channel estimation errors and imperfect interference cancellation, a novel NOMA-based power allocation scheme is proposed to meet diverse quality of service requirements.

\subsection{Contributions}
This study aims to create a HAPS system the serves terrestrial users within a $ 100 ~\rm{km}$ 
coverage area. It specifically focuses on tackling the challenge posed by the spatial correlation between the channel gain of multiple users and the antenna elements of the BS.
The spatial correlation among antenna elements, which we characterized through a spatial correlation matrix within the correlated Rician fading channel, is a critical parameter to consider. As the level of spatial correlation increases, the implications become apparent. In the case of an uncorrelated channel for the scenario with $M$ antenna elements on the BS, $M$ users experience orthogonal channels, corresponding to the degrees of freedom (DoF) of the channel. However, in scenarios where the channel is correlated, reflecting a correlation among antenna elements on the BS, the non-diagonal elements of the spatial correlation matrix tend to be non-zero. Consequently, the DoF of the spatial correlation matrix decreases, resulting in a reduction in the number of users with orthogonal channels. Consequently, the channels of users become more correlated, hindering the separation of signals using the detection matrix. This issue is particularly salient for HAPS BS due to several reasons: The high altitude of the HAPS BS leads to low angular resolvability in both elevation and azimuth angles. This inherent limitation results in increased correlation among antenna elements as their number rises.
Furthermore, the sparsity in the scattering environment between HAPS and terrestrial users results in a non-uniform contribution of the signal within the environment. Consequently, this leads to an increased correlation coefficient among the channels of user. Our paper is pioneering in its consideration of spatial correlation within the HAPS framework. Unlike previous literature, which did not account for spatial correlation in their analyses, our work addresses this crucial aspect, contributing to a more comprehensive understanding of the HAPS system. 
This paper's  main contributions are as follows: 
\begin{itemize}

\item \textbf{Sectorized antenna architecture:} Spatial correlation among antenna elements becomes more pronounced as the number of elements increases. To address this,  our proposal recommends employing multiple sectorized antenna arrays instead of a single large antenna array. This approach is proposed to mitigate the impact of spatial correlation on the overall performance of the system.
\item\textbf{Non-co-located polarization:} In this particular scenario, the polarization strategy for the antenna elements involves a systematic alternation between vertical and horizontal orientations in an interleaved fashion. For instance, odd-numbered elements are configured to transmit signals in the vertical polarization, whereas even-numbered elements transmit signals in the horizontal polarization. This alternating pattern ensures the orthogonality of signals for adjacent elements. This design choice results in a correlation of zero between signals transmitted by neighboring antenna elements. By employing this polarization interleaving and orthogonal signaling approach, the system aims to minimize interference and enhance the overall performance and reliability of signal reception.

\item \textbf{Grouping users into NOMA clusters:} Due to the LoS path between HAPS and terrestrial users, the spatial correlation between adjacent users is high. To address this problem, terrestrial users are grouped into NOMA clusters. In the downlink mode, since all users receive the same signal, the spatial correlation between them is not a problem. These NOMA clusters are formed based on two criteria: the strength of their correlation in LoS channel gain and their similarity in azimuth angle.

\item \textbf{Time slot allocation:}
The proposed time slot allocation algorithm addresses the challenge of interference among sectors, particularly in the case that the beamwidth of each sector is wide. The concern arises due to the potential overlap of signals from adjacent sectors. To counteract this interference, a strategic scheduling approach is introduced. This involves allocating distinct time slots to adjacent sectors, ensuring that their operations are temporally separated. By implementing this scheduling algorithm, the system can effectively manage and minimize interference, allowing each sector to transmit and receive signals without significant overlap. This temporal separation enhances the overall efficiency and reliability of the communication system, especially in scenarios where interference among adjacent sectors poses a potential hindrance to seamless operation.

\item \textbf{Maximizing energy efficiency and rate:}
In \cite{10008738}, the focus was on solving the power allocation problem to maximize the total rate while adhering to QoS and SIC conditions within our system. However, this paper extends the scope by addressing the challenge of maximizing energy efficiency alongside the primary goal of optimizing the total rate. This expanded objective reflects a comprehensive approach, considering both communication performance and energy conservation aspects in the system design. The dual emphasis on rate maximization and energy efficiency underscores the broader optimization goals undertaken in this specific research endeavor.

\item \textbf{Proving the optimality of the proposed power allocation algorithm:}
We provide proof the optimality of the proposed power allocation algorithm in Appendix C.
\end{itemize}

\begin{figure*}[htbp]
\begin{centering}
\centerline{\includegraphics[width=18cm]{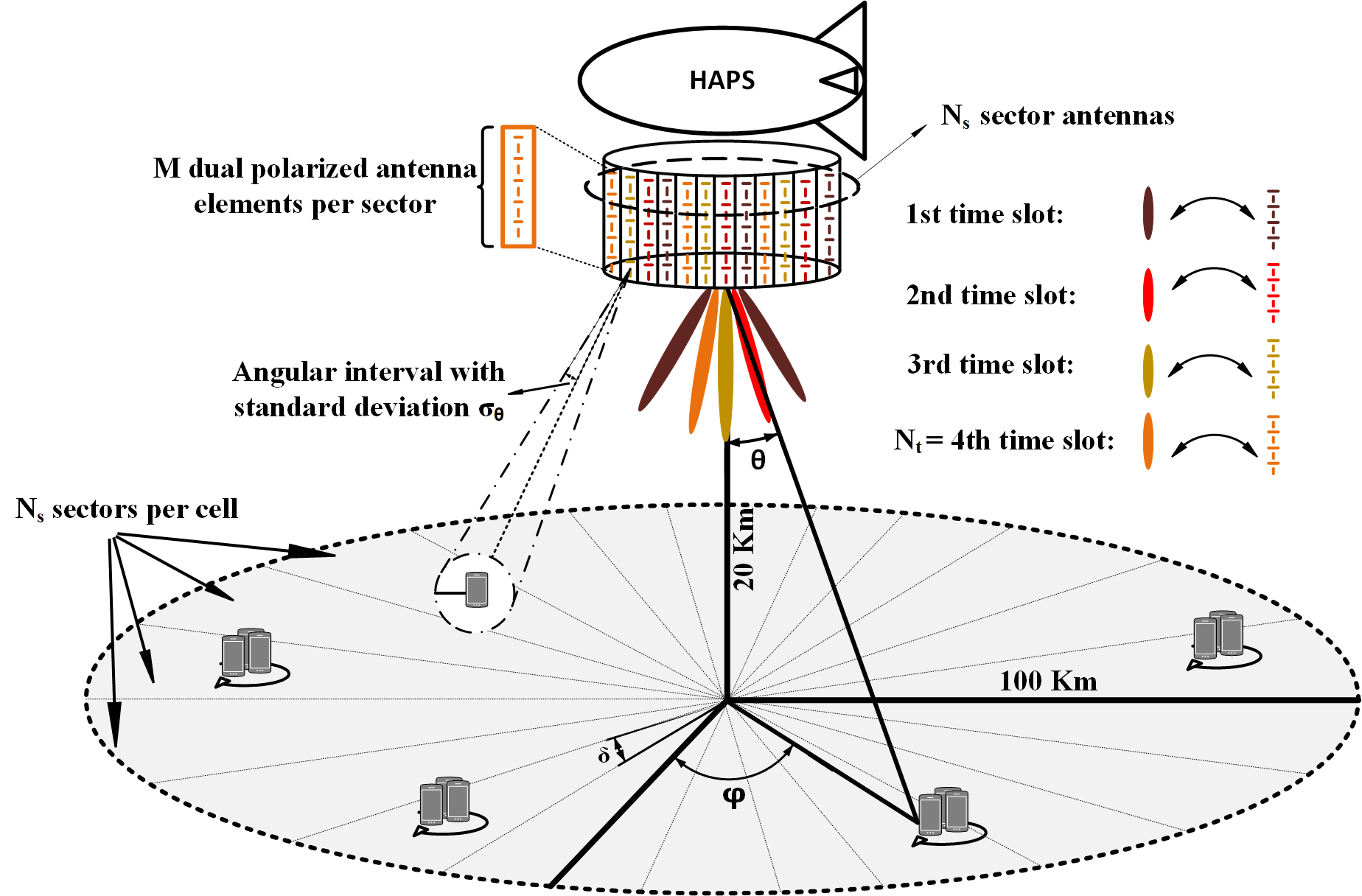}}
\caption{The proposed cylindrical antenna architecture with $N_s$ ULA sectors.}
\label{antennaModel}
\end{centering}
\end{figure*}

\section{System Model}
Fig. \ref{antennaModel} illustrates a communication network comprising terrestrial users and a HAPS BS. The HAPS is positioned at an altitude of $20~\mathrm{km}$, while the terrestrial users are distributed within a cell radius of $100~\mathrm{km}$. In the proposed system, which is based on MIMO NOMA technique, the aggregated signals for terrestrial users are transmitted within the sub-6 GHz frequency band.
The cylindrical antenna structure depicted in Fig. \ref{antennaModel} consists of vertical ULA antennas. Each ULA is responsible for transmitting the signals associated with NOMA clusters, ensuring that each terrestrial user is served by a single ULA. 
The elements of the vertical ULA antenna are positioned at $\left({m - 1}\right)d_v\lambda$ along the z-axis, where $0 < m \leq M$. Here, $\lambda$ denotes the wavelength of the signal, while $d_v$ represents the distance between adjacent antenna elements in the ULA array (in wavelengths), which consists of $M$ elements. The overall number of ULAs deployed in the system is denoted as $N_s$.

Additionally, Fig. \ref{antennaModel} provides specific information about the azimuth angle ($\varphi$) and elevation angle ($\theta$) assigned to a user within the network. These angular parameters play a crucial role in characterizing the spatial orientation of the user's signal reception and transmission.

\subsection{Channel model for ULA-based cylindrical antenna}
The HAPS BS is situated at a high elevation, resulting in few scatterers near it. Most scatterers are located around ground users, leading to a non-uniform distribution of signals in the environment. Due to the non-uniform distribution of the signal in the HAPS propagation environment, and the presence of an LoS path between the HAPS and users, the  channel for user $l$ in a  cluster $m$ is spatially correlated and a tractable method of modeling it is to utilize the Gaussian random variable with constant mean and spatial correlation matrix as shown below: 
\begin{equation}
{\bf h}_{m,l}^n  \sim {\rm N} \left( {{\bf \bar h}_{m,l}^n ,{\bf C}_{m,l}^n  }
\right).
\end{equation}
The term ${\bf \bar h}_{m,l}^n$ represents  the LoS component of ULA $n$ with $M$ antenna elements for the $l^{th}$  user and in the $m^{th}$ cluster, which is given by
\begin{equation}
{\bf \bar h}_{m,l}^n  = \sqrt {\beta _{m,l}^{\rm{LoS,n}} } \left[ {1,e^{j2\pi d_v \sin \left( \theta_{m,l}  \right)} , \cdots ,\left. {e^{j2\pi d_v \left( {M - 1} \right)\sin \left( {\theta _{m,l} } \right)} } \right]} \right.^T, 
 \end{equation}
 where  $\theta _{m,l}$ represents the angle at which the HAPS antenna is viewed from the position of user $(m,l)$, while $\beta_{m,l}^{\rm{LoS}}$ represents the large-scale fading for the LoS path, as given in \cite{10008738}.
The covariance matrix  ${\bf C}_{m,l}^n  $ characterizes the spatial correlation among the NLoS components.
 In other words, it describes the statistical relationship between the signal fading at different antenna positions\cite{8620255}, \cite{8445853}.
Referring to the local scattering model and the Gaussian angular distribution of $\theta$,  the $(a,b)^{th}$ entry of the spatial correlation matrix  (${{\bf C}_{m,l}^n } \in \mathbb{C}^{M \times M}$) is given by
\begin{equation}
\begin{array}{l}
\left[ {{\bf C}_{m,l}^n } \right]_{a,b}   = \beta _{m,l}^{\rm NLoS , n} \int_{ - \infty }^\infty  {e^{j2\pi d_v \left( {a - b} \right)\sin \bar \theta } f\left( {\bar \theta } \right)d\bar \theta }  = \\
\beta _{m,l}^{\rm NLoS , n} \int_{ - \infty }^\infty  {e^{j 2\pi d_v \left( {a - b} \right)\sin \left( {\theta  + \delta } \right)} \frac{1}{{\sqrt {2\pi } \sigma _\theta  }}e^{ - \frac{{\delta ^2 }}{{2\sigma _\theta ^2 }}} d\delta } ,
\end{array}
 \end{equation}
where   $\beta _{m,l}^{\rm NLoS, n} $ is the large-scale fading for NLoS path \cite{10008738}. ${\bar \theta } $ denotes the angle of an arbitrary multipath component as $ \bar \theta  = \theta  + \delta $, where $\theta $ is a deterministic nominal angle and $ \delta $ is a random deviation from the nominal angle with standard deviation $\sigma _\theta   = 5^ \circ$ \cite{SIG-093}.
The large-scale fading for the LoS and NLoS components in \rm{dB} are given by \cite{6863654}
\begin{equation}
\beta _{m,l}^{\rm LoS, n}  =  20\log d_{m,l}^n + 20\log f + 20\log \left( {\frac{{4\pi }}{c}} \right) + F_{m,l}^{\rm{LoS},n} ,
 \end{equation}
\begin{equation}
\beta _{m,l}^{\rm NLoS,n}  =  20\log d_{m,l}^n + 20\log f + 20\log \left( {\frac{{4\pi }}{c}} \right)  + F_{m,l}^{\rm{NLoS,n}} ,
 \end{equation}
where $F_{m,l}^n  \sim N\left( {0,\sigma _{{\rm sf}}^2 } \right) $ is the shadow fading with standard deviations $ \sigma _{{\rm sf}}  = 1 $ for LoS and $ \sigma _{{\rm sf}}  = 20 $ for NLoS links, and ${d_{m,l}^n }$ is
the distance between user $l$ in the cluster $m$ and HAPS in meters.

The likelihood of establishing a direct LoS connection between an antenna element on the HAPS and a terrestrial user relies on both the elevation angle and urban statistical characteristics. To quantify this probability, it can be described using the following formula \cite{6863654}, \cite{9838511}:
\begin{equation}
Pr\left( {{\rm LoS}} \right) = \frac{1}{{1 + \kappa \exp \left( { - \omega \left( {\theta - \kappa } \right)} \right)}},
\end{equation}
within this equation, $\kappa$ and $\omega$ are constants that vary depending on the characteristics of the environment.
Moreover, we can easily calculate the probability of NLoS path as $Pr\left( {{\rm NLoS}} \right) = 1 - Pr\left( {{\rm LoS}} \right)$  \cite{7881122}.

If the eigenvalue decomposition of  the spatial correlation matrix ${\bf C} \in \mathbb{C}^{M \times M}$ is given as ${\bf C} = {\bf UDU}^H$, the channel vector can be obtained using the Karhunen-Loeve expansion of ${\bf h}_{m,l}^n$ \cite{8861014} as 
\begin{equation}
{\bf h}_{m,l}^n = {\bf \bar h}_{m,l}^n  + ({\bf C}^n_{m,l})^{\frac{1}{2}} {\bf e} = {\bf \bar h}_{m,l}^n  + {\bf UD}^{\frac{1}{2}}  {\bf e},
\label{kkk}
 \end{equation}
where ${\bf e} \sim N\left( {{\bf 0}_M ,{\bf I}_M } \right)$.  
The equation \eqref{kkk} models the channel at a single antenna element of the user, simplifying the understanding of channel dynamics. Importantly, this model can be easily extended to accommodate users equipped with multiple antenna elements.  For instance, as described in section \rom{3}, where users have $N$ antenna elements, the channel gain between a specific ULA and a user is denoted by $\mathbf{H}_{m,l}^n$,  which is a matrix with dimensions $N \times M$. In this context, $N$ and $M$ represent the number of antenna elements for the users and the ULA, respectively.

 \subsection{Channel model with dual-polarized antennas}
When the spatial correlation between antenna elements becomes significant, especially with a high number of elements, non-co-located polarized antennas can provide a viable solution \cite{SIG-093}.
In this approach, the polarization of the antenna elements alternates between vertical and horizontal orientations in an interleaved manner. For example, the odd-numbered elements transmit their signals vertically, while the even-numbered elements transmit their signals horizontally.

\begin{figure}[htbp]
\centerline{\includegraphics[width=8cm]{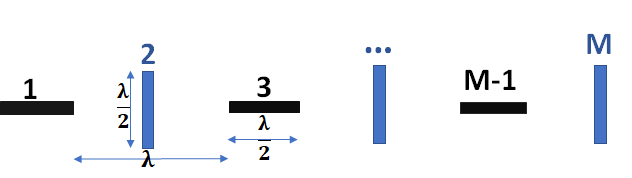}}
\caption{Dual-polarized non-co-located antenna \cite{SIG-093}.}
\label{polarized}
\end{figure}

By implementing non-co-located polarized antennas, it becomes possible to mitigate the spatial correlation between adjacent antenna elements. This is achieved by utilizing orthogonal polarizations, i.e., vertical and horizontal, which create two independent and parallel channels. In each channel, the distance between the antenna elements effectively doubles, resulting in reduced correlation.
Fig. \ref{polarized} illustrates a general model for a BS antenna array comprising M non-co-located antennas with dual polarization. The UE in this model is equipped with a uni-polarized antenna.
Let ${\bf h}_{(2i-1),V}$ and ${{\bf h}_{2i,H} }$ represent the channel coefficients from the UE to the $i^{th}$ vertically and horizontally polarized antenna, respectively. These coefficients are applicable for $i = 1, \cdots ,\frac{M}{2}$ that is given by
 \begin{equation}
{\bf h}_{m,l}^n  = \left[ {\begin{array}{*{20}c}
   { {\bf h}_{(2i-1),V} }  \\
   {{\bf h}_{2i,H} }  \\
\end{array}} \right].
 \end{equation}

\subsection{User clustering and time slot allocation }
\begin{algorithm}
  \caption{ User-clustering and time slot allocation. }
  \label{algorithm1}
  \textbf{User clustering:}\\
1. Derive the azimuth angle of the channel gain for each user.\\ 
2. Group users according to their azimuth angles: 
\begin{equation*}
\begin{array}{l} 
 \left( {n - 1} \right)\delta  < \varphi \left( n \right) < n\delta, \; \; \; \delta  = \frac{{2\pi }}{{N_s }},\; \;n = 1, \cdots ,N_s  .\\ 
 \end{array}
 \end{equation*}
3. Calculate the correlation coefficient between different users in each group according to the LoS channel gain as: \\
\(r_{i,j}  = \frac{{\left\| {{\bf \bar h}_i {\bf \bar h}_j^t } \right\|}}{{\left\| {{\bf \bar h}_i } \right\|\left\| {{\bf \bar h}_j } \right\|}}\).\\
4. if  \(r_{i,j}  \ge \rho  \Rightarrow  \) 
put user $i$ and user $j$  in the same cluster.\\
\\
\textbf{Time slot allocation:}\\
1. Initialization:\;   $\phi \left( \varphi  \right) , \delta  $ \\ 
2. Calculate  the number of time slots:\\
$ \phi \left( \varphi  \right) =  - 12\left( {\frac{\varphi }{{\varphi _{3dB} }}} \right)^2 dB
 $\cite{9779715},\; \; $  N_t  = 2\left\lceil {\frac{{\phi \left( \varphi  \right)  - \delta }}{{2\delta }}} \right\rceil +1. $ \\
3. Assign different time slots to adjacent sectors: \\
$\begin{array}{l}
 {\rm 1st \; time \; slot} = \left\{ {\varphi \left( 1 \right),\varphi \left( {1 + N_t } \right),\varphi \left( {1 + 2N_t } \right), \cdots } \right\}, \\ 
 {\rm 2nd \; time \; slot} = \left\{ {\varphi \left( 2 \right),\varphi \left( {2 + N_t } \right),\varphi \left( {2 + 2N_t } \right), \cdots } \right\}, \\ 
 {\rm k - th \; time \; slot} = \left\{ {\varphi \left( k \right),\varphi \left( {k + N_t } \right),\varphi \left( {k + 2N_t } \right), \cdots } \right\}. \\ 
 \end{array}$
\end{algorithm}
In the downlink mode of communication, it is observed that users within the same NOMA cluster experience identical signal reception. Consequently, even in scenarios where a high spatial correlation exists among the channel gains of users within a cluster, there is no detrimental impact on signal propagation \cite{10008738}. Each NOMA cluster is exclusively served by a single sector utilizing a ULA configuration. The clustering process initially involves grouping users based on their azimuth angles, followed by calculating correlation coefficients among the channel gains of multiple users. Finally, users exhibiting high spatial correlation are allocated to the same cluster.
\begin{figure}[htbp]
\centerline{\includegraphics[width =10cm]{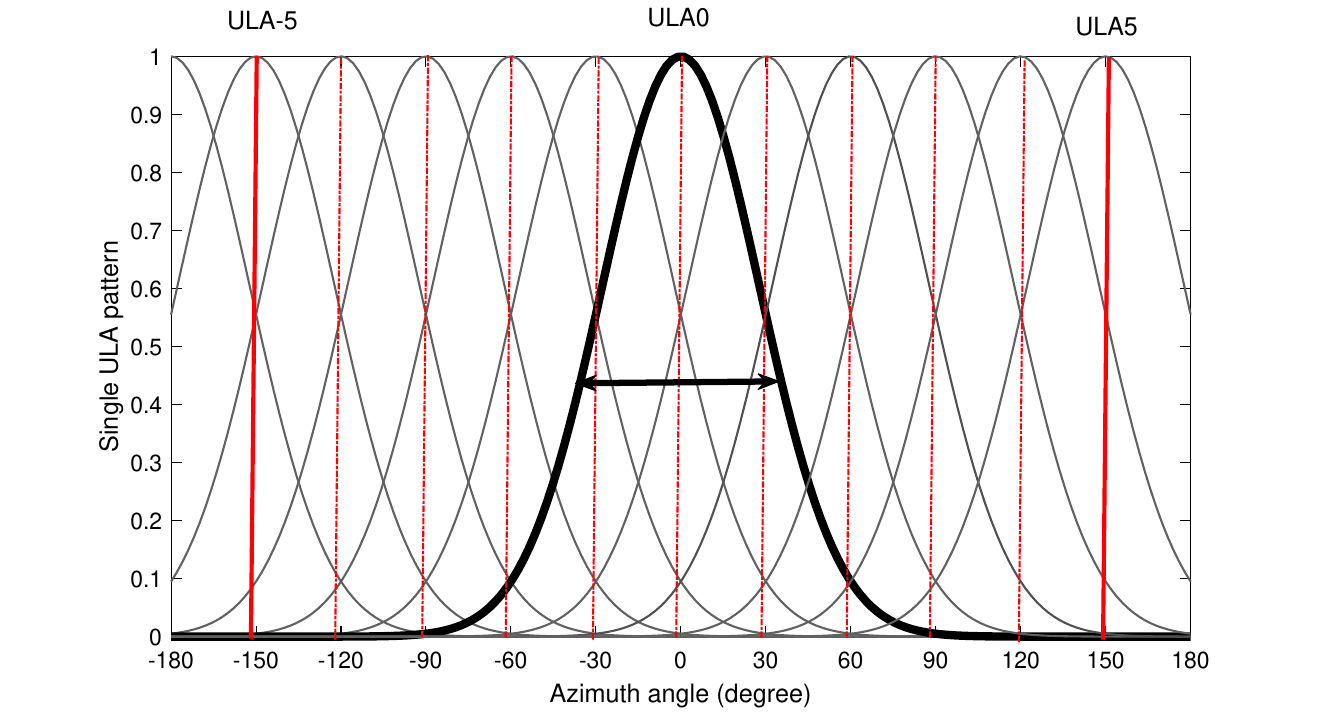}}
\caption{Beam pattern for adjacent ULAs.}
\label{ULAUPA}
\end{figure}
\begin{figure}[htbp]
\centerline{\includegraphics[width=6cm]{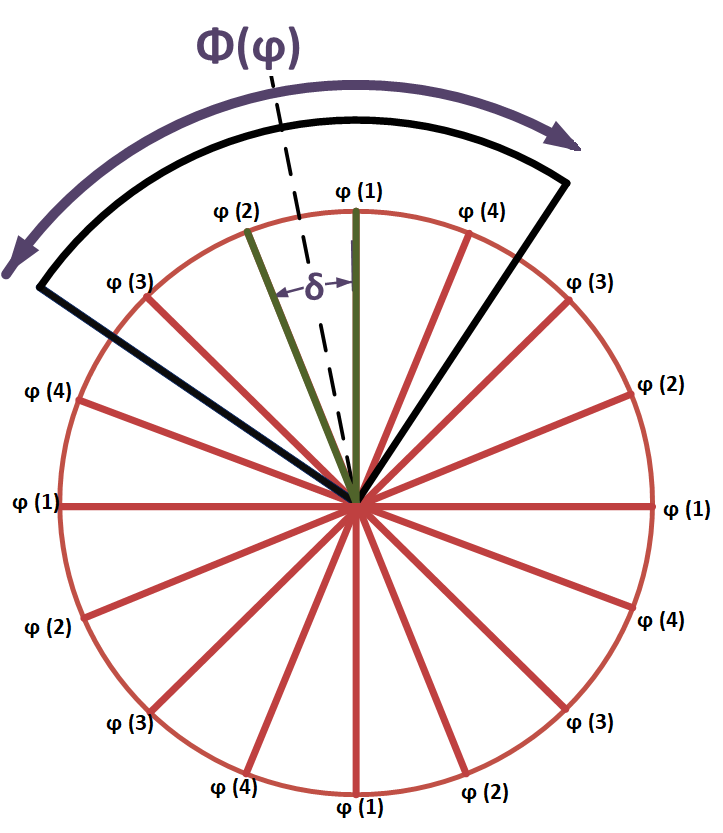}}
\caption{The number of interfering ULAs per each ULA.}
\label{beamwidth}
\end{figure}
Interference arising from adjacent sectors can be a concern, particularly when the beamwidth of each sector is high. To mitigate such interference, a scheduling approach is employed where adjacent sectors are allocated separate time slots. In order to determine the appropriate number of time slots required within the proposed scenario, the beamwidth equation must be considered \cite{9779715}. Algorithm 1 is introduced as a solution to allocate these time slots between sectors, effectively preventing interference. In an effort to maximize frequency utilization, all sectors within the system share the same frequency bandwidth.
 To accurately determine the radiation power pattern of each sector, we refer to the beam pattern model specified in \cite{9779715}. Initially,  we ascertain the beamwidth of each element in the azimuth domain by $\phi \left( \varphi  \right) =  - 12\left( {\frac{\varphi }{{\varphi _{3dB} }}} \right)^2 $. 
As depicted in Fig. \ref{ULAUPA}, the beamwidth of each sector in the azimuth domain exhibits interference with adjacent sectors.
Referring to Fig. \ref{beamwidth}, in order to quantify the number of sectors that overlap with the selected sector, we employ the equation $N_t = 2\left\lceil \frac{{\phi(\varphi) - \delta}}{{2\delta}} \right\rceil + 1$. Here, $\phi(\varphi)$ represents the radiation pattern in decibels $\rm{(dB)}$, $ \varphi_{3dB}$ denotes the horizontal $3~\rm{dB}$ beamwidth, and $\delta$ corresponds to the azimuth angle of each sector. Subsequently, we allocate distinct time slots to the interfering sectors in order to mitigate interferences among them.
 
\section{Signal Processing}

The corresponding transmitted signals from the HAPS sectors are modulated by a precoding matrix $ {\bf P} \in \mathbb{C}^{M \times M}  $ and then transmitted over the radio channel.
The precoding matrix  used by each of the sector antenna in downlink mode is  indicated by $ {\bf P} \in \mathbb{C}^{M \times M}$, which is equal to the identity matrix ($ {\bf P} = I_{_{M}}$) \cite{7236924,8301007}. 
Therefore, the transmitted  signals of each sector  can be expressed as
\begin{equation}
 \begin{aligned}
{\bf x}^n  = {\bf Ps}^n,
\label{eq3}
\end{aligned}
\end{equation}
where the $ M \times 1 $ vector $ \bf s $ is given by \cite{8301007}
\begin{equation}
\begin{aligned}
{\bf s}^n  =\left[ {\begin{array}{*{20}c}
   {\sqrt {P_{\max } \Omega _{1,1}^n } s_{1,1}^n   +  \cdots  + \sqrt {P_{\max } \Omega _{1,L}^n } s_{1,L}^n }  \\
    \vdots   \\
   {\sqrt {P_{\max } \Omega _{M,1}^n } s_{M,1}^n  +  \cdots  + \sqrt {P_{\max } \Omega _{M,L}^n } s_{M,L}^n }  \\
\end{array}} \right].
\label{eq4}
\end{aligned}
\end{equation}
Each element of the above matrix represents a superposed signal transmitted to the $ m^{th} $ cluster that is served by  $ n^{th} $ sector.
$P_{max}$ denotes the maximum power of the HAPS BS for serving MIMO NOMA users.
 $\Omega _{m,l}^n $ denotes the NOMA power allocation coefficient for user $ (m,l) $ served by sector $n$, and  $s _{m,l}^n $ is the symbol of the  $ l^{th} $ user  in the $ m^{th} $ cluster.
The received signal for the $ l^{th} $ user of the $ m^{th} $ cluster can be expressed as follows:
 \begin{equation}
\begin{aligned}
{\mathbf{ y}}_{m,l}^n  = {\mathbf H}_{m,l}^n {\bf Ps}^n + {\mathbf n}_{m,l}^n,
\label{eq5}
\end{aligned}
\end{equation}
where $ {\mathbf n}_{m,l}^n $ is the additive white Gaussian  noise with variance $ \sigma ^2$. 
 To further examine the user's observed signal, we can express it as follows:
\begin{equation}
{\mathbf{ y}}_{m,l}^n  = {\bf H}_{m,l}^n {\bf p}_m \sum\limits_{l = 1}^L {\sqrt {P_{\max } \Omega _{m,l}^n } } { s}_{m,l}^n  + \sum\limits_{k = 1,k \ne m}^M {{\bf H}_{m,l}^n {\bf p}_k {\bf s}_k }  + {\mathbf n}_{m,l}^n .
 \label{eq6}
\end{equation}

The detection vector for user $ (m,l) $ is denoted as $ {\bf v}_{m,l}^n$. To completely mitigate inter-cluster interference, the precoding and detection matrices must meet the condition $ \left( {{\mathbf v}_{m,l}^n } \right)^H {\mathbf H}_{m,l}^n {\mathbf p}_k = 0 $, for all $k \ne m $, where $ {\bf p}_{k} $ stands for the $ k ^{th}$ column of $ {\bf P} $. 
In our investigation of MIMO-NOMA systems, we  assumed perfect channel state information (CSI) for the design of detection vectors \( \mathbf{v}_{m,l}^n \) \cite{7236924}, \cite{8301007}.
After implementing the detection vector $ {\bf v}_{m,l} $, the received signal can be reformulated as follows:
\begin{equation}
\begin{array}{l}
\left( {{\mathbf v}_{m,l}^n } \right)^H  {\mathbf y}_{m,l}^n  = \left( {{\mathbf v}_{m,l}^n } \right)^H  {\mathbf H}_{m,l}^n {\mathbf p}_m \sum\limits_{l = 1}^L {\sqrt {P_{\max } \Omega _{m,l}^n } s_{m,l}^n } 
\\ \; \; \; \; \; \; \; \; \; \; \; \; \; \; \;   \; \; \; \; \; + \left( {{\mathbf v}_{m,l}^n } \right)^H  {\mathbf n}_{m,l}^n.
\label{eq7}
\end{array}
\end{equation}
By applying the SIC to the received signal of each user, the interference is eliminated  from the users with lower channel gain. According to appendix \ref{appendix1-outage}, the following conditions must be satisfied:
\begin{equation}
P_{\max } \Omega _{m,l}^n \gamma _{m,l - 1}^n  - P_{\max } \sum\limits_{k = 1}^{l - 1} {\Omega _{m,k}^n \gamma _{m,l - 1}^n }  \ge P_{tol} ,\; \; l = 2,3, \cdots ,L,
\end{equation}
where $ P_{\rm{tol}} $ is the minimum power difference required to differentiate the signal to be decoded and the remaining non-decoded signals\cite{7557079}. 
The value of $ \gamma _{m,l - 1}^n$ is defined as effective channel gain and equals $\left| {\left( {{\mathbf v}_{m,l-1}^n } \right)^H  {\mathbf H}_{m,l-1}^n {\mathbf p}_m } \right|^2 $. Users in each NOMA cluster are sorted according as
\begin{equation}
\left| {\left( {{\mathbf v}_{m,1}^n } \right)^H  {\mathbf H}_{m,1}^n {\mathbf p}_m } \right|^2  \ge  \cdots  \ge \left| {\left( {{\mathbf v}_{m,L}^n } \right)^H  {\mathbf H}_{m,L}^n {\mathbf p}_m } \right|^2.
\end{equation}
Therefore,  the user $ (m,l) $ sees interference from only users with greater channel gain than itself, and the data rate of user $ (m,l) $ is given by
\begin{equation}
\begin{aligned}
{R_{m,l}^n}   = \log _2 \left( {1 + \frac{{\rho \Omega _{m,l}^n \left| {\left( {{\mathbf v}_{m,l}^n } \right)^H  {\mathbf H}_{m,l}^n {\mathbf p}_m } \right|^2}}{{1 + \rho \sum\nolimits_{k = 1}^{l - 1} {\Omega _{m,k}^n\left| {\left( {{\mathbf v}_{m,l}^n } \right)^H  {\mathbf H}_{m,l}^n {\mathbf p}_m } \right|^2} }}} \right),
\label{eq10}
 \end{aligned}
 \end{equation}
where $ \rho  = \frac{{P_{\max } }}{{\sigma ^2 }} $.

\section{Problem Formulation}
The goal of this paper is to find the power allocation coefficients of MIMO terrestrial users to maximize energy efficiency while ensuring that each user achieves a pre-defined minimum data rate. The problem can be formulated as follows:
\begin{maxi!}|s|[2]
	{\Omega_{m,l}^n }{ {\eta _{EE}} \label{eq:ObjectiveExample3}}
	{\label{eq:Example3}}
    {}
\addConstraint{R_{m,l}^n  \ge R^{\rm{QoS}} ,m \in \left\{ {1, \cdots ,M } \right\},l \in \left\{ {1, \cdots ,L } \right\}\label{eq:constr-1}}
\addConstraint{\begin{array}{l}
 P_{\max } \Omega _{m,l}^n \gamma _{m,l - 1}^n  - P_{\max } \sum\limits_{k = 1}^{l - 1} {\Omega _{m,k}^n } \gamma _{m,l - 1}^n  \ge P_{\rm{tol}}\\ 
 l = 2,3, \cdots ,L, \\ 
 \end{array} 
 \label{eq:constr-2}}
\addConstraint { {\sum\limits_{n = 1}^{N_s } {\sum\limits_{m = 1}^M {\sum\limits_{l = 1}^L {\Omega _{m,l}^n } } }  \le 1  } \label{eq:constr-3}}
\addConstraint {
\Omega _{m,l}^n  \ge 0, \; \; \; \forall m,l,n.
\label{eq:constr-4}}
\end{maxi!}
The energy efficiency, denoted by $ \eta _{EE} $, is a metric defined by the formula $ \eta _{EE}  = \frac{{R^{sum} }}{{P_t }}$
, where $R^{sum} $ represents the total throughput and $ P_t$ denotes the total power consumption of HAPS.
Constraints \eqref{eq:constr-1}  and \eqref{eq:constr-2}  ensure the QoS and SIC conditions of users, respectively, while \eqref{eq:constr-3} limits the power of the HAPS system. 
 $ P_{\rm{max}} $ denotes the maximum power of the HAPS BS encompassing all users and clusters.
The objective function of equation \eqref{eq:ObjectiveExample3} is non-convex, and hence obtaining an optimal solution for this problem is challenging. 
To address this problem effectively, we initially approach it by focusing on maximizing the corresponding spectral efficiency problem, which is more manageable \cite{8301007}.
The SE maximization problem can be formulated as
\begin{maxi!}|s|[2]
	{\Omega_{m,l}^n }{ {\sum\limits_{n = 1}^{N_s } {\sum\limits_{m = 1}^M {\sum\limits_{l = 1}^L {R_{m,l}^n } } } } \label{ObjectiveExample3}}
	{\label{Example3}}
    {}
\addConstraint{R_{m,l}^n  \ge R^{\rm{QoS}} ,m \in \left\{ {1, \cdots ,M } \right\},l \in \left\{ {1, \cdots ,L } \right\}\label{constr-1}}
\addConstraint{\begin{array}{l}
 P_{\max } \Omega _{m,l}^n \gamma _{m,l - 1}^n  - P_{\max } \sum\limits_{k = 1}^{l - 1} {\Omega _{m,k}^n } \gamma _{m,l - 1}^n  \ge P_{\rm{tol}}\\ 
 l = 2,3, \cdots ,L, \\ 
 \end{array} 
 \label{constr-2}}
\addConstraint { {\sum\limits_{n = 1}^{N_s } {\sum\limits_{m = 1}^M {\sum\limits_{l = 1}^L {\Omega _{m,l}^n } } }  \le 1  } \label{constr-3}}
\addConstraint {
\Omega _{m,l}^n  \ge 0, \; \; \; \forall m,l,n.
\label{constr-4}}
\end{maxi!}

This paper presents an iterative algorithm to tackle this issue.
 \begin{algorithm}
  \caption{ The power allocation algorithm.}
\textbf{1. Initialization.} \\
 $  R_{m,l}^{\min } ,P_{\rm{tol}} ,\rho \left| {\left( {{\bf v}_{m,l}^n } \right)^H {\bf H}_{m,l}^n {\bf p}_m } \right|^2 ,l \in \left\{ {1, \cdots ,L} \right\}.$
\\ \\
\textbf{2. Primary power allocation to satisfy the QoS and SIC constraints for all of users.}\\
$  \Omega _{m,l}^{\rm{QoS}}  = \left( {2^{R_{m,l}^{\rm{QoS}} }  - 1} \right)\left( {\sum\limits_{k = 1}^{l-1} {\Omega _{m,k}^{\min } }  + \frac{1}{{\rho {\left| {\left( {{\mathbf v}_{m,l}^n } \right)^H  {\mathbf H}_{m,l}^n {\mathbf p}_m } \right|^2} }}} \right),  $\\
$ \Omega _{m,l}^{\rm{SIC}}  = \sum\limits_{k = 1}^{l-1} {\Omega _{m,k}^{\min } }  + \frac{{ P_{\rm{tol}}  }}{{ \rho {\left| {\left( {{\mathbf v}_{m,l-1}^n } \right)^H  {\mathbf H}_{m,l-1}^n {\mathbf p}_m } \right|^2} }} \; \; for \; \;l\geq 2.$
\\
Satisfy both conditions simultaneously $\Rightarrow$\\ $ \Omega _{m,l}^{\min }  = \max \left\{ {\Omega _{m,l}^{\rm{QoS}} ,\Omega _{m,l}^{\rm{SIC}} } \right\}.  $ \\ \\
\textbf{3. Allocation of residual power of HAPS BS between clusters and users.}
\\ 
a- Allocation of power among the clusters:

Arrange the clusters according to their fractional level.$\Rightarrow$ $ \left[ {H,m} \right] = {\rm{sort}}\left( {\frac{{P_{\max } 2^{\sum\nolimits_{l = 1}^L {R_{m,l}^{\min } } } }}{{\rho \left| {\left( {{\mathbf v}_{m,1}^n } \right)^H  {\mathbf H}_{m,1}^n {\mathbf p}_m } \right|^2}}} \right)$.\\
b- Allocation of power among users in the cluster:\\
    ${\bf while} \; ({\rm P}_{{\rm rem}}  > 0):$ \\ $\begin{array}{l}
 \left\{ {\Omega _{m\left( i \right),1}^{n,\min } } \right. +  = \frac{{H\left( {i + 1} \right) - H\left( i \right)}}{{P_{\max } }}, \\ 
 \left. {\rm{update}\left( {\Omega _{m\left( i \right),j}^{n,\rm{QoS}} ,\Omega _{m\left( i \right),j}^{n,\rm{SIC}} ,\Omega _{m\left( i \right),j}^{n,\min } } \right),i +  = 1,j = 2:L} \right\} \\ 
 \end{array}$.
\\  \\
\textbf{4. Calculate:}\\
$R_{m\left( i \right),j}^{\min }  = \log _2 \left( {1 + \frac{{\rho \Omega _{m\left( i \right),j}^{\min } \left| {{\bf v}_{m\left( i \right),j}^H {\bf H}_{m\left( i \right),j} {\bf p}_{m\left( i \right)} } \right|^2 }}{{1 + \rho \sum\nolimits_{k = 1}^{j - 1} {\Omega _{m\left( i \right),k}^{\min } \left| {{\bf v}_{m\left( i \right),j}^H {\bf H}_{m\left( i \right),j} {\bf p}_{m\left( i \right)} } \right|^2 } }}} \right) \;$.
\label{}
\end{algorithm}
In order to adhere to the constraints of SIC and QoS, the initial phase of the power allocation algorithm involves assigning a minimum amount of power to each user. Subsequently, the additional power available from the HAPS can be distributed among both users and clusters to maximize the total system data rate. To achieve this goal, the power allocation process is carried out in two stages: initially, power distribution is determined for each cluster, followed by the distribution of power among the clusters themselves.

In \cite{8301007}, the authors allocated the power of a BS among users and clusters to maximize the total rate of the system.
In their proposed algorithm, the power of the BS was first allocated among all users in the cell to satisfy the QoS condition, and then the remaining power was allocated to the user of each cluster with the highest channel gain. This is because these users do not receive interference from other users, and they also have the best channel gain to maximize the total rate.
However, the user of each cluster with the best channel gain causes interference for other users in the cluster, and for this reason, allocating the remaining power to this user violates the QoS and SIC constraints for other users in the cluster.
 
 We obtain the minimum power allocation coefficients of MIMO-NOMA users to satisfy the QoS and SIC constraints in  \eqref{eq:constr-1} and \eqref{eq:constr-2} as follows:
\begin{equation}
 \Omega _{_{m,l} }^{\rm{QoS} }  = \left( {2{}^{R_{_{m,l} }^{\rm{QoS }} } - 1} \right)\left( {\sum\nolimits_{k = 1}^{l - 1} {\Omega _{_{m,k} }  + \frac{1}{\rho \left| {\left( {{\mathbf v}_{m,l}^n } \right)^H  {\mathbf H}_{m,l}^n {\mathbf p}_m } \right|^2}} } \right) ,
\end{equation}
\begin{equation}
\Omega _{m,l}^{\rm{SIC}}  = \sum\limits_{k = 1}^{l - 1} {\Omega _{m,k}  + \frac{{P_{\rm{tol}} }}{{\rho \left| {\left( {{\mathbf v}_{m,l-1}^n } \right)^H  {\mathbf H}_{m,l-1}^n {\mathbf p}_m } \right|^2 }}} .
\end{equation}
According to \eqref{eq:constr-3}, the power of the HAPS is limited, and in order to satisfy the QoS and SIC constraints of users,  the HAPS must have  a minimum amount of power, which is obtained from the following equation:
\begin{equation}
P_{\rm{req}}  = P_{\max } {\sum\limits_{n = 1}^{N_s } {\sum\limits_{m = 1}^M {\sum\limits_{l = 1}^L {\Omega _{m,l}^{n,\min} } } }},
\end{equation}
where  $ \Omega _{m,l}^{n, \min }  $ is the minimum power allocation for each user that is generated by the power allocation algorithm.
To satisfy the SIC and QoS constraints, we allocate minimum power to all users in the first part of the power allocation algorithm. In this way, the additional power of the HAPS can be allocated among users and clusters in order to maximize the overall system rate.
To this end, we first determine the power allocation within each cluster, and then we distribute power among clusters.

\begin{proposition}\label{proposition-outage}
Within a NOMA cluster, the user with the highest channel gain maximizes the overall system rate better than any other user.
\end{proposition}
\begin{proof}
See Appendix \ref{appendix-outage}.
\end{proof}
The user who has the strongest channel gain performs better in maximizing the overall rate of the system compared to other users in the cluster. This is because this user is not experiencing interference from other users and has a stronger channel gain compared to the others.
Nevertheless, this user's transmissions can cause interference to other users in the same cluster. Therefore, allocating additional power to the user with the highest channel gain can result in a violation of the QoS and SIC constraints of other users. Therefore, it is necessary to update the QoS and SIC constraints of all other users in the same cluster when assigning the extra power to users with the strongest channel gain. In order to distribute power among clusters, we utilize the following equation \cite{8301007}:
\begin{equation}
\Delta P_m^n  = \left( {2^{\Delta R}  - 1} \right)\frac{{P_{\max } 2^{\sum\nolimits_{l = 1}^L  {\hat R_{m,l}^n  } } }}{{\rho \left| {\left( {{\bf v}_{m,1}^n } \right)^H {\bf H}_{m,1}^n {\bf p}_m } \right|^2 }},
 \end{equation}
where $ \Delta P_m^n $ refers to the extra power needed to increase the sum rate of the $m^{th}$ cluster in the $n^{th}$ sector by $ \Delta R$.
The notation $\hat R_{m,l}^n$ is used to signify the rate achieved by user $ (m,l)$ within sector $n$.
It is evident that if $ \Delta P_m^n $ remains constant, a smaller value of  $ \frac{{P_{\max } 2^{\sum\nolimits_{l = 1}^L {\hat R_{m,l}^n  } } }}{{\rho \left| {\left( {{\bf v}_{m,1}^n } \right)^H {\bf H}_{m,1}^n {\bf p}_m } \right|^2 }} $ yields a higher data rate. This notion can be used as a foundation for the power allocation algorithm between clusters. Initially, we calculate the fraction amount of power for each cluster, which is given by  $ \frac{{P_{\max } 2^{\sum\nolimits_{l = 1}^L {\hat R_{m,l}^n  } } }}{{\rho \left| {\left( {{\bf v}_{m,1}^n } \right)^H {\bf H}_{m,1}^n {\bf p}_m } \right|^2 }} $.
Then, the fractional levels are sorted in ascending order, beginning with the lowest level. Power is subsequently distributed to each fractional level until it reaches the next one. Next, power is allocated to the first two fractions until they equal the third fraction, and this operation is repeated until the HAPS's extra power is exhausted.
It is important to note that the fraction  $ \frac{{P_{\max } 2^{\sum\nolimits_{l = 1}^L {\hat R_{m,l}^n  } } }}{{\rho \left| {\left( {{\bf v}_{m,1}^n } \right)^H {\bf H}_{m,1}^n {\bf p}_m } \right|^2 }} $ is obtained by assigning power to the first user in the $m^{th}$ cluster and $n^{th}$ sector. If power is added to only the first user in the cluster, it violates the QoS and SIC requirements for the other users. As previously discussed, power is allocated to the other users in the cluster as long as their QoS and SIC conditions are met.
The procedure for allocating the power of HAPS BS to maximize the total rate subject to the QoS and SIC conditions of the users is outlined in Algorithm 2.

In the context of the water-filling algorithm, there is a final water level determined by the power limitation of the BS. Clusters with fraction levels below the final level receive power to reach the final level, while clusters with higher fraction levels do not receive additional power.
To establish the ultimate level, we incorporate a supplementary variable $ \lambda$. When $ \lambda  \le \frac{{P_{\max } 2^{\sum\nolimits_{l = 1}^L {R_{m,l}^{\min } } } }}{{\rho \left| {{\bf v}_{m,1}^H {\bf H}_{m,1} {\bf p}_m } \right|^2 }}$, no power is allocated to the $m^{th}$ cluster, and it remains unaffected. However, if $ \lambda  > \frac{{P_{\max } 2^{\sum\nolimits_{l = 1}^L {R_{m,l}^{\min } } } }}{{\rho \left| {{\bf v}_{m,1}^H {\bf H}_{m,1} {\bf p}_m } \right|^2 }}$, the $m^{th}$ cluster receives supplementary power to attain the desired final level.

Energy efficiency is a crucial aspect in the design and operation of HAPS BS. The energy efficiency of a HAPS BS can be defined as the ratio of the total rate achieved to the power consumed. 
Therefore, we can derive the expression for energy efficiency using only the variable $\lambda$ as
\begin{equation}
\eta _{EE}  = \frac{{\sum\limits_{m = 1}^M {\sum\limits_{l = 1}^L {R_{m,l}^{\min } } }  + \sum\limits_{m = 1}^M {\left[ {\log _2 \left( \lambda  \right) - \log _2 \left( {\frac{{P_{\max } 2^{\sum\nolimits_{l = 1}^L {R_{m,l}^{\min } } } }}{{\rho \left| {{\bf v}_{m,1}^H {\bf H}_{m,1} {\bf p}_m } \right|^2 }}} \right)} \right]^ +  } }}{{P_{req}  + \sum\limits_{m = 1}^M {\left[ {\lambda  - \frac{{P_{\max } 2^{\sum\nolimits_{l = 1}^L {R_{m,l}^{\min } } } }}{{\rho \left| {{\bf v}_{m,1}^H {\bf H}_{m,1} {\bf p}_m } \right|^2 }}} \right]^ +  } }}.
\label{energy eff}
\end{equation}
In the given equation, $\left[ x \right]^+$ denotes the positive value of a quantity, equivalent to $x$ when $x$ is positive or zero.
Mathematically, the energy efficiency equation has the piecewise form of the logarithm of power in the numerator divided by an affine function of power in the denominator. The logarithmic nature of the numerator and the affine nature of the denominator result in the energy efficiency function being piecewise concave. At lower power levels, an increase in power consumption yields improvements in both the total rate and energy efficiency. This behavior arises from the fact that boosting power at lower levels enhances signal quality, leading to higher data rates and improved energy efficiency. Consequently, the total rate and energy efficiency exhibit a positive correlation within this range.
However, as power levels increase beyond a certain threshold, further increments in power consumption yield diminishing improvements in the total rate relative to the increase in power. Consequently, the energy efficiency starts to decline, as the logarithmic function in the numerator exhibits slower growth compared to the increase in power in the denominator.
For lower power levels, the pursuit of maximum energy efficiency coincides with the objective of maximizing the total rate. This alignment arises due to the positive correlation between power, total rate, and energy efficiency within this regime. However, as power levels escalate and energy efficiency begins to diminish, the goals of maximizing the total rate and maximizing energy efficiency diverge. In this scenario, maximizing the total rate might involve compromising energy efficiency due to the diminishing returns associated with higher power consumption.
The maximum value of energy efficiency is obtained by finding the unique root of the equation $ \frac{{\partial \eta _{EE} }}{{\partial \lambda }} = 0 $. After reaching this specific power level, any further increase in power leads to higher data rates but diminished energy efficiency.
Indeed, when the power is below the threshold, maximizing energy efficiency aligns with increasing data rate. However, beyond the threshold, maximizing energy efficiency no longer corresponds to maximizing data rate.

\begin{proposition}\label{proposition2-outage}
 This algorithm allocates power between users optimally.
\end{proposition}
\begin{proof}
See Appendix \ref{appendix2-outage}.
\end{proof}
We proved the optimality of our power allocation algorithm. Consequently,
the algorithm reliably converges towards the optimal solution. This convergence underscores the effectiveness of our proposed approach in iteratively refining the power allocation until a state is reached that
corresponds to an optimal solution.
The complexity of the proposed power allocation algorithm is $ O\left( {M \times L \times \left( {L - 1} \right)} \right)$, which is the number of operations to satisfy the QoS and SIC conditions. Here, $M$ refers to the number of antenna elements, while $L$ represents the number of users per cluster.

\section{Simulation Results}
This section presents simulation results for the proposed scenario to demonstrate its performance gains.  
The simulation parameters are summarized in Table I. 
\begin{centering}
\begin{table}[h] 
\centering 
\begin{tabular}{c c } 
 \hline\hline 
\textbf{Table 1:} Simulation parameters. \\ 
  \hline \hline
   The area radius  &  $ 100\; \rm{km} $  \\ 
Carrier frequency $(f_c)$  &  $ 2.5 \; \rm{GHz} $  \\
 Bandwidth &  $ 10\; \rm{MHz}$   \\  
 Elevation of the BS &  $ 20\; \rm{km}$  \\ 
  $R_{\min}$ (minimum rate of each user) &  $ 1\; {{\rm{bps}} \mathord{\left/
 {\vphantom {{\rm{bps}} {\rm{Hz}}}} \right.
 \kern-\nulldelimiterspace} {\rm{Hz}}} $  \\ 
   The power spectral density of thermal noise & $  - 174 \; \rm{dBm/Hz}  $ \\
    Noise figure & $ 7\; \rm{dB}$ \\
   $ \varphi _{3dB}$  & $ 65^ \circ $ \\
  $ P_{\rm{tol}} $ & $ 10\; \rm{dBm} $ \\
  $ \left( {\sigma _{\rm{SF}}^{\rm{LoS}} ,\sigma _{\rm{SF}}^{\rm{NLoS}} } \right) $ & $(4,6) \, \rm{dB} $ \cite{3gpp.36.33}\\[1ex] 
    \hline 
    \end{tabular} \label{table:nonl} 
    \end{table}
\end{centering}
We investigate the spectral efficiency and the energy efficiency of ULA cylindrical HAPS in different scenarios. 
 The spatial correlation, which is influenced by the elevation angle, decreases the rank of the channel matrix and subsequently degrades the quality of signal detection. Note that in order to have fair comparisons, we assume that the same amount of time-frequency resources are utilized at all simulation.
 
\begin{figure}[htbp]
\centerline{\includegraphics[width =8cm]{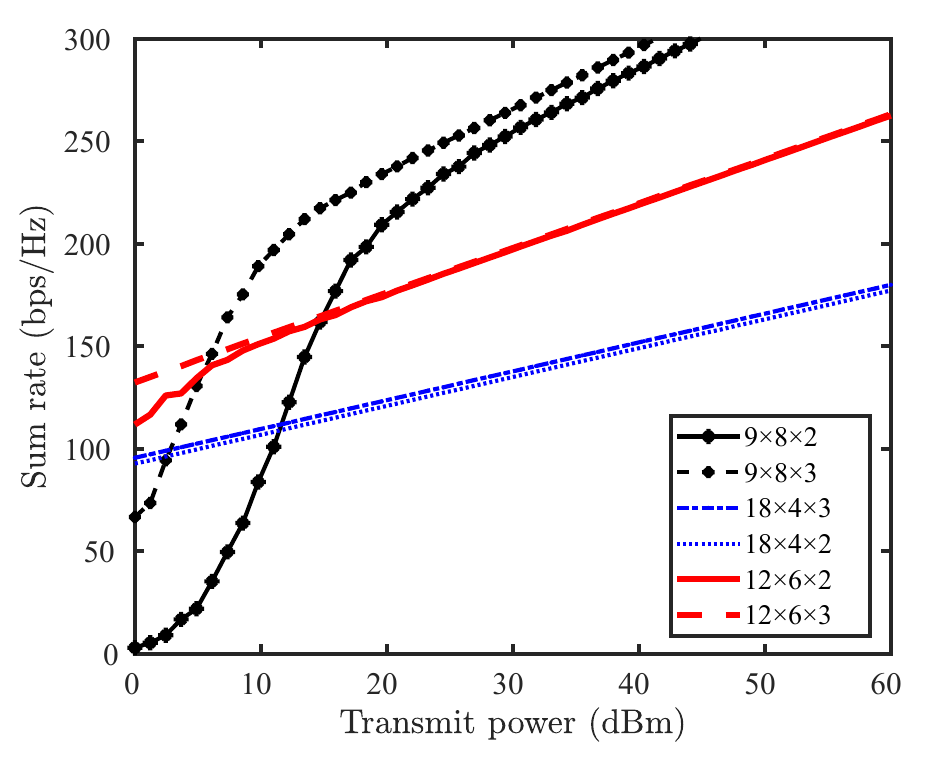}}
\caption{ Total sum rate versus the power of HAPS BS for a fixed total number of elements $ (M=72)$. The minimum rate to satisfy the QoS is $(R_{\rm{min}}=1 \; \rm{bps/Hz}) $  and antenna spacing equals $2\lambda$. The ULA scenario $ \left( {18 \times 4 \times 2} \right)$ consists of $18$ ULA sectors, with $4$ clusters and $2$ users in each cluster.} 
\label{totalRate}
\end{figure}

 Fig. \ref{totalRate} shows the sum rate of a massive MIMO NOMA system for ULA cylindrical antennas with a constant total number of antenna elements and different elements per sector.
Two or three users are assigned to each NOMA cluster in each scenario. For instance, ULA $ \left( {18 \times 4 \times 2} \right)$   indicates that there are $18$ sectors, each comprising $4$ elements, and capable of serving $4$ clusters with two users in each cluster. 
While the spatial correlation between antenna elements in a  ULA cylindrical antenna  with $8$ elements per sector is greater than that of a  ULA with $4$  and 6 elements,  the spectral efficiency of ULA $ \left( {9 \times 8 \times 2} \right)$ is higher than that of  ULA $ \left( {18 \times 4 \times 2} \right)$ and $ \left( {12 \times 6 \times 2} \right)$,  because the former has fewer sectors, and therefore consumes fewer resource blocks.
Note that for lower power, the available signal strength is inadequate for the effective separation of user signals using the detector. Consequently, the ULA sector with $8$ antenna elements exhibits lower performance compared to other configurations due to its higher spatial correlation. Consequently, the detector encounters challenges when operating at lower power levels with the ULA with $8$ antenna elements. 
Furthermore, Fig. \ref{totalRate} illustrates the gap between two cases that there are two and three users per cluster, particularly evident at low power levels. This gap becomes more pronounced as the number of elements per sector increases. The elevated correlation and interference associated with a high number of elements impede the detection process, leading to a concentration of power allocation through the proposed primary power allocation to meet QoS and SIC conditions.
Consequently, it is evident that three users per cluster yield a higher data rate compared to scenarios involving two users, emphasizing the impact of correlation and interference on the effectiveness of signal separation and power allocation strategies.

\begin{figure}[htbp]
\centerline{\includegraphics[width =9cm]{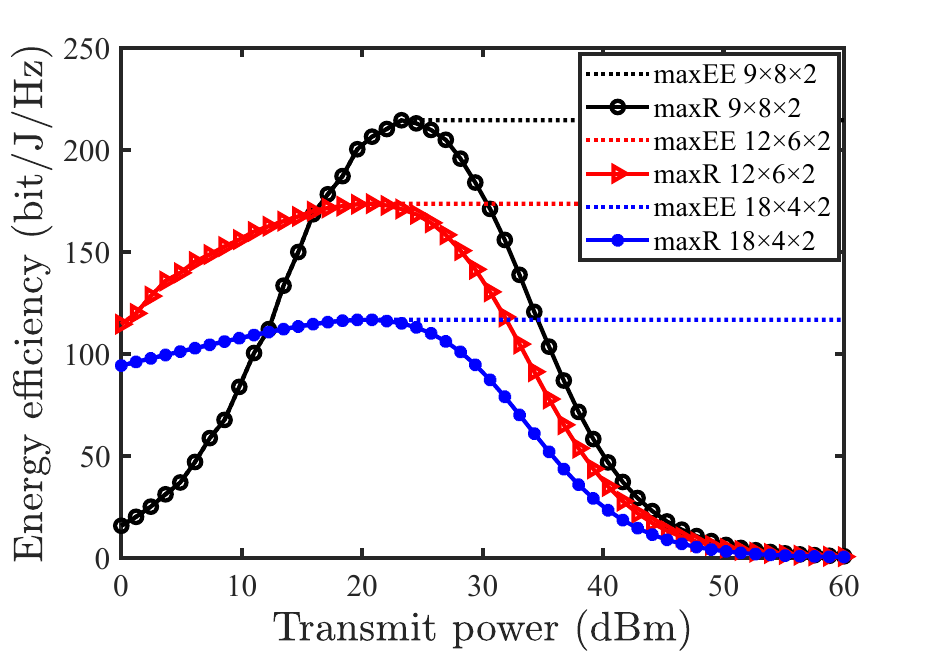}}
\caption{Trade-off between energy efficiency and HAPS BS power for sum rate maximization and energy efficiency maximization scenarios. The minimum QoS rate is set to $(R_{\rm{min}}=1 \; \rm{bps/Hz})$, and the antenna spacing is fixed at $2\lambda$.}
\label{EER}
\end{figure}

Energy efficiency varies concavely with power, and it reaches a single maximum point determined by finding the sole root of the equation $ \frac{{\partial \eta _{EE} }}{{\partial \lambda }} = 0 $. Once this critical power level is attained, further increasing power leads to higher data rates but reduced energy efficiency. Consequently, the energy efficiency with maximization energy efficiency remains constant for power levels beyond this threshold, and the energy efficiency associated with the maximum sum rate decreases. As illustrated in Fig. \ref{EER}, when power is below the threshold, maximizing energy efficiency (maxEE) aligns with increasing data rates (maxR). However, beyond the threshold, the maximum energy efficiency remains constant. As a result, in the subsequent figures, only the energy efficiency for the total rate maximization scenario is indicated.

\begin{figure}[htbp]
\centerline{\includegraphics[width = 9cm ]{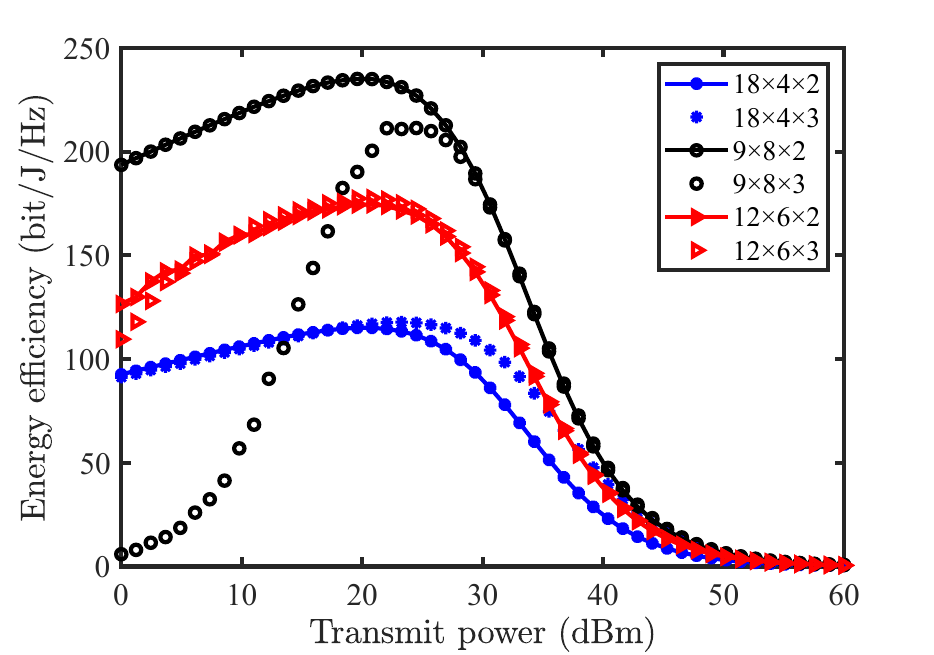}}
\caption{Energy efficiency versus the power of HAPS BS. The total number of elements equals $72$ and the distance between antenna elements equals $2\lambda$. The ULA scenario $ \left( {18 \times 4 \times 2} \right)$ consists of 18 ULA sectors, with 4 clusters and 2 users in each cluster.}
\label{EE1}
\end{figure}

Fig. \ref{EE1}, compares the energy efficiency of the proposed ULA  cylindrical antennas.
For lower transmit powers, the three-user model is less energy efficient than the two-user model. However, this trend reverses when the power of HAPS BS increases. When there are fewer users in the cluster, the QoS and SIC conditions can be more effectively met with a lower power from the HAPS BS, and the extra power can be used to maximize the data rate.
However,  when there is enough power, the QoS and SIC conditions for users can be met easily and the total rate for three users in the cluster exceeds that of two users. The spatial correlation between antenna elements in a ULA cylindrical antenna with more elements per sector is greater compared to a ULA with fewer elements. However, the energy efficiency of the ULA with fewer sectors (and consequently fewer utilized resource blocks) is higher than the ULA with more sectors.
A discernible gap is observed across all scenarios between the cases with two and three users per cluster, as depicted in Fig. 8. This gap is particularly pronounced with an increasing number of elements per sector. 
Raising the number of elements per sector induces spatial correlation among antenna elements, consequently amplifying interference among users. 
Specifically, for low power, the detection vector exhibits limitations, exacerbating the issue as the number of users per cluster increases from 2 to 3.
This limitation is conspicuous across all scenarios for two and three users. Notably, the gap diminishes for scenarios involving 4 elements per sector, i.e., $18\times4\times2$ and $18\times4\times3$, where the correlation and interference are substantially lower. However, for subsequent scenarios, i.e., $12\times6\times2$ and $12\times6\times3$, the gap intensifies, reaching a peak for the scenarios $9\times8\times2$ and $9\times8\times3$ due to heightened correlation and interference.

\begin{figure}[htbp]
\centerline{\includegraphics[width =8cm]{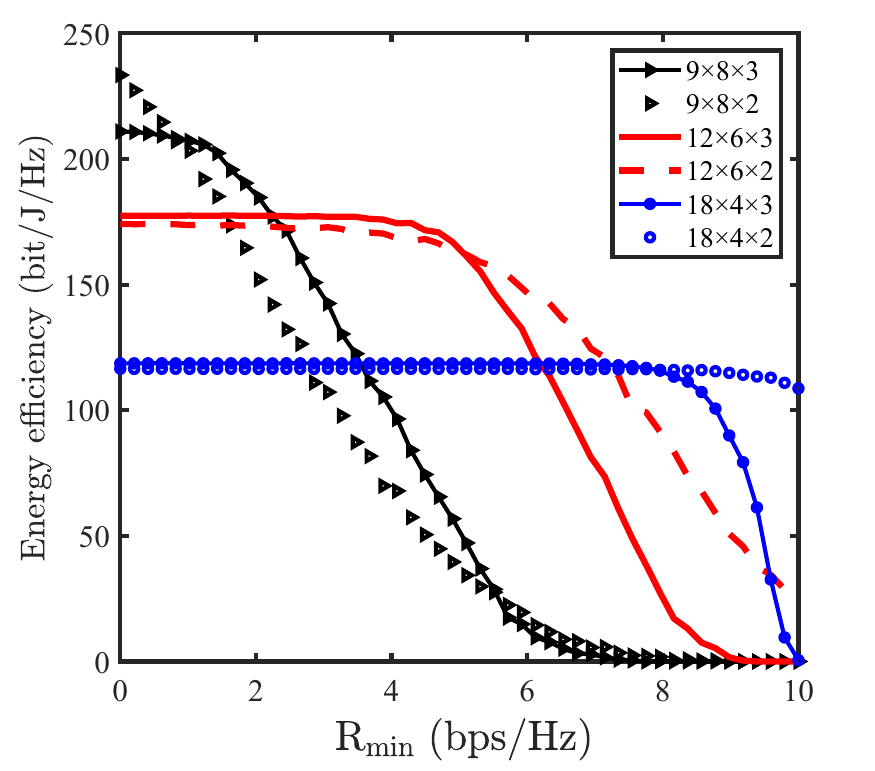}}
\caption{ Comparison of energy efficiency versus the minimum rate of user $ (R_{\rm{min}})$ with
$ 72$ elements.}
\label{EERmin}
\end{figure}
In Fig. \ref{EERmin}, the relationship between energy efficiency and the minimum rate requirement of each user is depicted. 
As the minimum rate requirement for QoS rises, the HAPS BS must allocate more power to meet the higher data rate demands of the users. However, due to its restricted power capacity, allocating more power to satisfy the increasing rate demands reduces the available power for other system operations. Consequently, the HAPS BS's potential to fulfill QoS constraints for all users diminishes as the minimum rate requirement increases.
At first, as the QoS minimum rate constraint is increased, the required power also increases, while the energy efficiency remains relatively constant. However, when the minimum rate is increased too significantly, the BS may not have enough power to satisfy the QoS requirements for all users, resulting in a decrease in energy efficiency.
Among various scenarios, the ULA configuration with 8 antenna elements shows higher energy efficiency for lower $R_{\rm{min}}$ values compared to other setups, mainly due to its more effective resource allocation. However, for higher $R_{\rm{min}}$ values, its performance is adversely affected because of increased interference.

\begin{figure}[htbp]
\centerline{\includegraphics[width =9cm]{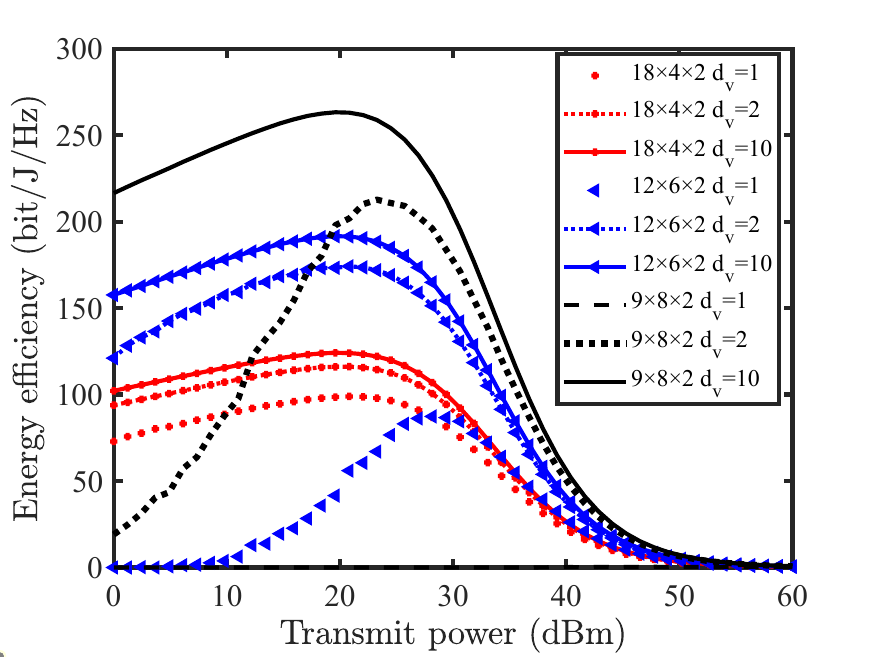}}
\caption{ Comparison of energy efficiency versus power for HAPS BS with
$ 72$ elements and $ 144$ users for different antenna spacing scenarios.}
\label{antennaSpace}
\end{figure}
The spatial correlation matrix is influenced by various factors such as the distance between antenna elements. By using a distance between antenna elements greater than $ \lambda /2$, the angular resolution of the array with a fixed number of elements is enhanced because the spatial correlation between antenna elements decreases.  
Fig. \ref{antennaSpace} illustrates the impact of the distance between antenna elements on energy efficiency. As shown in Fig. \ref{antennaSpace}, for different scenarios  increasing the distance between the antenna elements leads to improved energy efficiency.
When the distance between antenna elements on the BS is small compared to the wavelength of the transmitted signal, the signals received at the adjacent antennas are highly correlated. This is because the signals arrive at the antennas with only a small phase difference, which is a function of the distance between the antennas and the direction of the signal's arrival. Consequently, the correlation between adjacent antennas is high, and the signals can be treated as almost identical.
\begin{figure}[htbp]
\centerline{\includegraphics[width =9cm]{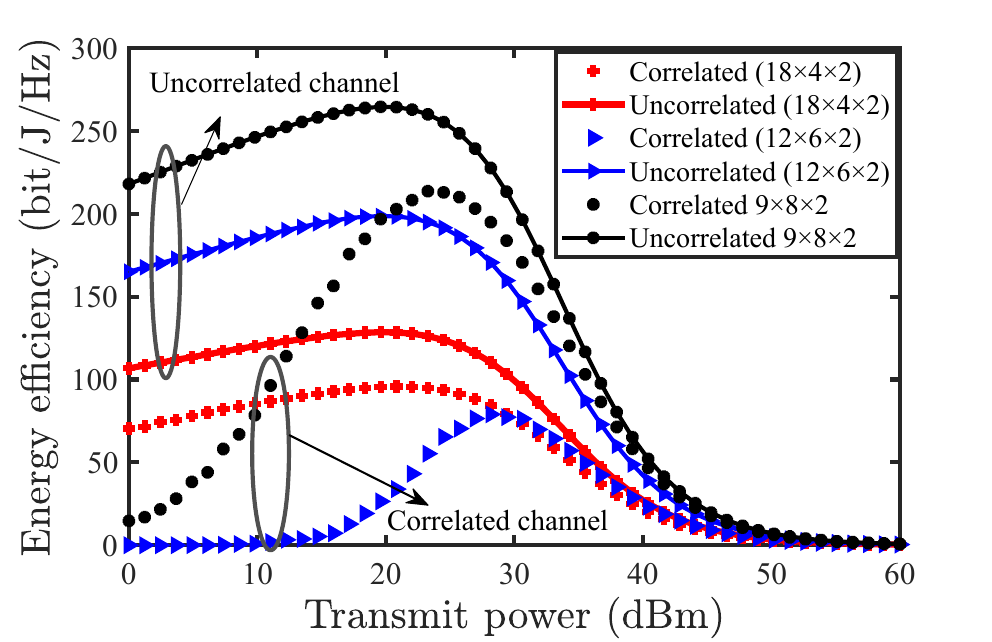}}
\caption{Energy efficiency versus the power of HAPS BS, with the total number of elements $ M=72$ and $144$ users for correlated and uncorrelated channels.}
\label{withoutR}
\end{figure}
\par
 Fig. \ref{withoutR}, illustrates the difference between spatially uncorrelated and correlated channels. 
Spatial correlation creates large eigenvalue variations, and the correlation matrix contains dominant spatial eigenvectors.
Furthermore, signals propagate in specific directions when the spatial correlation is high.  
As the spatial correlation level increases, the rank of the channel matrix decreases and signal detection becomes difficult. 
However, in the uncorrelated channel case, the correlation matrix is the identity matrix and is a full rank matrix,  so it has no spatial directivity. 
For this reason, the energy efficiency of the uncorrelated channel is higher than that of  the correlated channel. 
 \begin{figure}[htbp]
\centerline{\includegraphics[width =9cm]{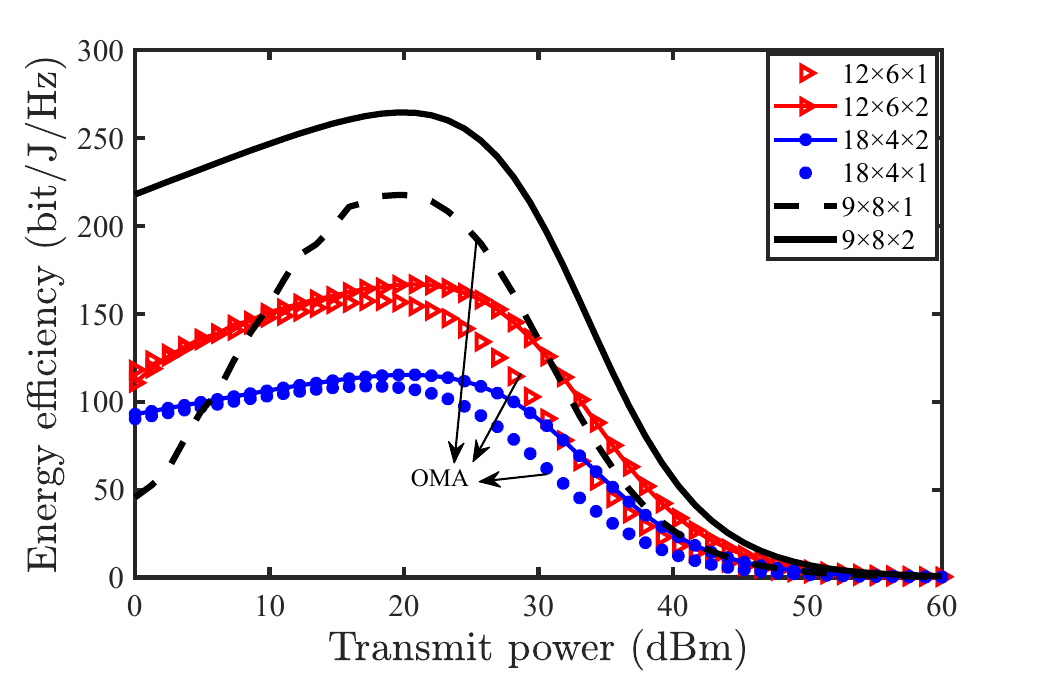}}
\caption{Comparison of energy efficiency between OMA and NOMA with $ 72$ elements. }
\label{aaa}
\end{figure}

Fig. \ref{aaa} compares the energy efficiency of NOMA and OMA for different scenarios. The comparison is conducted under the constraint that the total number of users, the number of antenna elements, and the utilized time-frequency resources remain constant for both techniques. NOMA enables multiple users to share the same frequency band, which increases spectral efficiency and subsequently improves energy efficiency. 
\begin{figure}[htbp]
\centerline{\includegraphics[width =9cm]{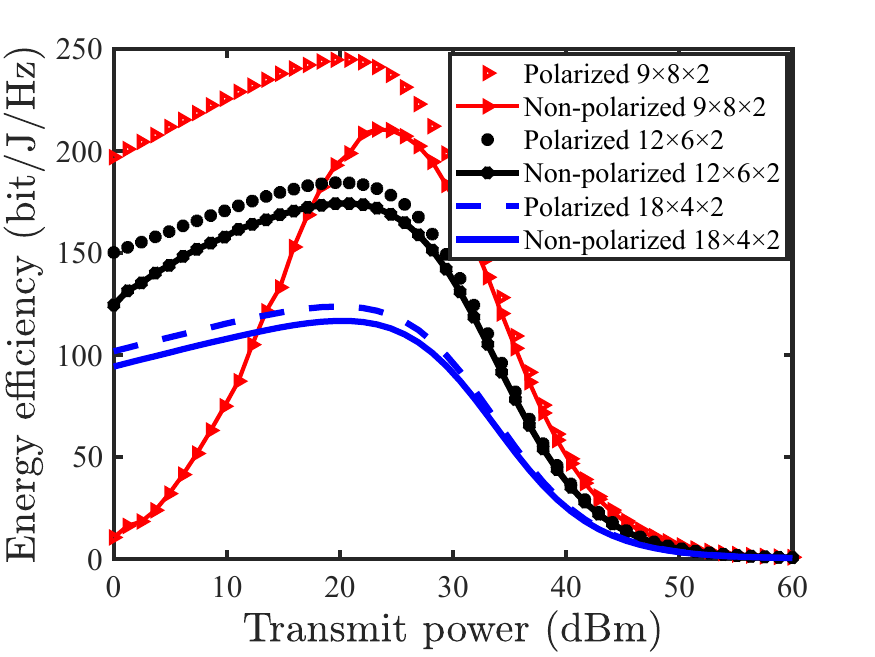}}
\caption{Comparison of energy efficiency versus power for HAPS BS with $ 72$ elements and $ 144$ users in both polarized and non-polarized scenarios.}
\label{CompareR}
\end{figure}

In Fig. \ref{CompareR}, the performance of a system using dual polarized and uni-polarized antennas is presented. 
When dual-polarized antennas are not co-located, they can be modeled as two separate antennas transmitting and receiving signals in orthogonal polarization states. In this case, the behavior of the dual-polarized antennas can be similar to using two parallel uni-polarized antennas, with each antenna experiencing a different propagation environment. This can result in a lower correlation between the channel gains of the two antennas.
The spatial correlation matrix is affected by the distance between antenna elements, and doubling the distance between the elements can result in a significant decrease in the spatial correlation. 
In Fig. \ref{CompareR}, it is evident that the energy efficiency difference between polarized and non-polarized antennas for ULA with 8 antenna elements is greater than the energy efficiency difference observed in ULAs with 6 and 4 antenna elements. This is due to the impact of polarization on the spatial correlation, which becomes more pronounced with an increased number of antenna elements.
In other words, as the number of antenna elements increases in the ULA, the spatial correlation between polarized and non-polarized antennas becomes more significant. This leads to a more substantial difference in energy efficiency between the two types of antennas in the ULA with 8 elements. On the other hand, the difference in energy efficiency between polarized and non-polarized antennas is less noticeable in the ULAs with 6 and 4 elements, where the impact of polarization on spatial correlation is comparatively smaller.

\section{conclusion}
This paper presents a novel approach for developing a  MIMO HAPS system that serves terrestrial users with spatially correlated channels. The authors proposed a cylindrical antenna architecture for the HAPS BS and consider the spatial correlation between the multi-path channel gain for each user and the spatial correlation among users. To address the issue of high spatial correlation among users, the paper proposed a NOMA clustering method to mitigate the impact of the spatial correlation matrix.   Next, a power allocation algorithm was presented to maximize the system's total rate while respecting the QoS and SIC conditions of users. Simulation results revealed that the spatial correlation significantly impacts both energy efficiency and spectral efficiency in  MIMO HAPS systems.  Comparing the spectral and energy efficiency for the two scenarios provides valuable insights for optimizing the system design.

\appendices

\section{ }
\label{appendix1-outage}
A proper selection of the transmission power for each NOMA user is required to carry out SIC.
As a condition for efficient SIC at the receiver of UE1, power allocation needs to satisfy the following:
\begin{equation}
P_{\max } \Omega _{m,3}^n \gamma _{m,1}  - P_{\max } \sum\limits_{k = 1}^2 {\Omega _{m,k}^n \gamma _{m,1}  }  \ge P_{tol},
\label{a}
\end{equation}
\begin{equation}
P_{\max } \Omega _{m,2}^n\gamma _{m,1}   - P_{\max } \Omega _{m,1}^n \gamma _{m,1}   \ge P_{tol}.
\label{b}
\end{equation}
Therefore, the condition for allocating power to eliminate the interference from UE3 on the UE2 receiver can be expressed as the following:
\begin{equation}
P_{\max } \Omega _{m,3}^n \gamma _{m,2}   - P_{\max } \sum\limits_{k = 1}^2 {\Omega _{m,k}^n  \gamma _{m,2}  }  \ge P_{tol}. 
\label{c}
\end{equation}
Note that since $ \gamma _1  \ge \gamma _2 $, (\ref{a}) automatically holds if (\ref{c})  is true. Therefore, in a NOMA cluster with $L$ users, the power constraints for efficient SIC are as follows:
\begin{equation}
P_{\max } \Omega _{m,l}^n \gamma _{m,l - 1}^n  - P_{\max } \sum\limits_{k = 1}^{l - 1} {\Omega _{m,k}^n \gamma _{m,l - 1}^n }  \ge P_{tol} ,l = 2,3, \cdots ,L.
\end{equation}

\section{Proof of Proposition \ref{proposition-outage}}
\label{appendix-outage}
To support the claim that the first user in the NOMA cluster contributes more to the total rate than the other users in the cluster, it has been demonstrated that allocating additional power from any user to the first user in the cluster results in an improved sum rate.  Assume that extra power $ \Delta P_{tr} $ is transferred from user $r$ to user $1$. The rates of users with weaker channel gains than the $r^{th}$ user remain unchanged since the total interference remains constant. The total rate of the initial $r$ users before  power transferring is as
\begin{equation}
\begin{array}{l}
 \sum\limits_{l = 1}^r {R_{m,l}^n }  = \sum\limits_{l = 1}^r {\log _2 \left( {\frac{{1 + \rho \sum\nolimits_{k = 1}^l {\Omega _{m,k}^n \gamma _{m,l}^n } }}{{1 + \rho \sum\nolimits_{k = 1}^{l - 1} {\Omega _{m,k}^n \gamma _{m,l}^n } }}} \right)}  = \\
 \log _2 \left( {\prod\limits_{l = 1}^{r - 1} {\frac{{1 + \rho \sum\nolimits_{k = 1}^l {\Omega _{m,k}^n \gamma _{m,l}^n } }}{{1 + \rho \sum\nolimits_{k = 1}^l {\Omega _{m,k}^n \gamma _{m,l + 1}^n } }} \times \left( {1 + \rho \sum\limits_{k = 1}^r {\Omega _{m,k}^n \gamma _{m,r}^n } } \right)} } \right). \\ 
 \end{array}
\end{equation}
The total rate of the first $r$ users after the power transfer is as
\begin{equation}
\begin{array}{l}
\sum\limits_{l = 1}^r {\left( {R_{m,l}^n } \right)^\prime  }  = \\
\log _2 \left( {\prod\limits_{l = 1}^{r - 1} {\frac{{1 + \rho \left( {\Delta P_{tr}  + \sum\nolimits_{k = 1}^l {\Omega _{m,k}^n } } \right)\gamma _{m,l}^n }}{{1 + \rho \left( {\Delta P_{tr}  + \sum\nolimits_{k = 1}^l {\Omega _{m,k}^n } } \right)\gamma _{m,l + 1}^n }} \times \left( {1 + \rho \sum\limits_{k = 1}^r {\Omega _{m,k}^n \gamma _{m,r}^n } } \right)} } \right).
 \end{array}
\end{equation}
In addition, we have this condition:
\begin{equation}
\begin{array}{l}
\gamma _{m,l}^n  \ge \gamma _{m,l + 1}^n  \Rightarrow \\ \frac{{1 + \rho \sum\nolimits_{k = 1}^l {\Omega _{m,k}^n \gamma _{m,l}^n } }}{{1 + \rho \sum\nolimits_{k = 1}^l {\Omega _{m,k}^n \gamma _{m,l + 1}^n } }} \le  
\frac{{1 + \rho \left( {\Delta P_{tr}  + \sum\nolimits_{k = 1}^l {\Omega _{m,k}^n } } \right)\gamma _{m,l}^n }}{{1 + \rho \left( {\Delta P_{tr}  + \sum\nolimits_{k = 1}^l {\Omega _{m,k}^n } } \right)\gamma _{m,l + 1}^n }}.
\end{array}
\end{equation}
Consequently, we can prove $ \sum\nolimits_{l = 1}^r {R_{m,l} }  < \sum\nolimits_{l = 1}^r {R'_{m,l} }$.

\section{Proof of Proposition \ref{proposition2-outage} }\label{appendix2-outage}
To prove the optimality of algorithm 2, assume two clusters $q$ and $n$. With the algorithm, each cluster receives extra power $ \Delta P_q $ and $ \Delta P_n $, and their sum rate increases with $ \Delta R_q $ and $ \Delta R_n $, respectively, as
\begin{equation}
\begin{array}{l}
 \Delta R_q  = \log _2 \left( {\Delta P_q  + \frac{{P_{\max } 2^{\sum\nolimits_{l = 1}^L {R_{q,l}^{\min } } } }}{{\rho \gamma _{q,1} }}} \right) - \log _2 \left( {\frac{{P_{\max } 2^{\sum\nolimits_{l = 1}^L {R_{q,l}^{\min } } } }}{{\rho \gamma _{q,1} }}} \right), \\ 
 \end{array}
\end{equation}
\begin{equation}
\begin{array}{l}
 \Delta R_n  = \log _2 \left( {\Delta P_n  + \frac{{P_{\max } 2^{\sum\nolimits_{l = 1}^L {R_{n,l}^{\min } } } }}{{\rho \gamma _{n,1} }}} \right) - \log _2 \left( {\frac{{P_{\max } 2^{\sum\nolimits_{l = 1}^L {R_{n,l}^{\min } } } }}{{\rho \gamma _{n,1} }}} \right), \\ 
 \end{array}
\end{equation}
\begin{equation}
\begin{array}{l}
 \Delta R_q ^\prime   = \log _2 \left( {\Delta P_q  + \Delta p + \frac{{P_{\max } 2^{\sum\nolimits_{l = 1}^L {R_{q,l}^{\min } } } }}{{\rho \gamma _{q,1} }}} \right) - \log _2 \left( {\frac{{P_{\max } 2^{\sum\nolimits_{l = 1}^L {R_{q,l}^{\min } } } }}{{\rho \gamma _{q,1} }}} \right), 
 \end{array}
\end{equation}
\begin{equation}
\begin{array}{l}
 \Delta R_n ^\prime   = \log _2 \left( {\Delta P_n  - \Delta p + \frac{{P_{\max } 2^{\sum\nolimits_{l = 1}^L {R_{n,l}^{\min } } } }}{{\rho \gamma _{n,1} }}} \right) - \log _2 \left( {\frac{{P_{\max } 2^{\sum\nolimits_{l = 1}^L {R_{n,l}^{\min } } } }}{{\rho \gamma _{n,1} }}} \right). 
 \end{array}
\end{equation}
The sum rate difference between two cases can be calculated as follows:
\begin{equation}
\begin{array}{l}
 \Delta R_{sum}  = \Delta R'_q  + \Delta R'_n  - \Delta R_q  - \Delta R_n  =  \\ 
 \log _2 \left( {\left( {\Delta P_q  + \frac{{P_{\max } 2^{\sum\nolimits_{l = 1}^L {R_{q,l}^{\min } } } }}{{\rho \gamma _{q,1} }}} \right)^2  - \left( {\Delta p} \right)^2 } \right) - \\
 \log _2 \left( {\Delta P_q  + \frac{{P_{\max } 2^{\sum\nolimits_{l = 1}^L {R_{q,l}^{\min } } } }}{{\rho \gamma _{q,1} }}} \right)^2  \le 0. \\ 
 \end{array}
\end{equation}
It is clear from the equation above that power shifting between two clusters reduces the sum rate.  This confirms the optimality of the algorithm.

\ifCLASSOPTIONcaptionsoff
  \newpage
\fi
\bibliographystyle{IEEEtran}
\bibliography{IEEEabrv,Bibliography}
\vfill
\end{document}